\documentclass[aps,prl,reprint,longbibliography,floatfix]{revtex4-2}

\usepackage[T1]{fontenc}
\usepackage{lmodern}
\usepackage{microtype}
\usepackage{amsmath,amssymb}
\usepackage{graphicx}
\usepackage{hyperref}
\hypersetup{hidelinks}

\newcommand{\qstar}{q_*}
\newcommand{\Iwin}{\mathcal I}

\begin{document}

\title{Constituent-Tagged Transfer Linking Cluster Dynamics to Static Scaling}

\author{Kenji Shinoda}
\email{kenji_shinoda@mail.toyota.co.jp}
\affiliation{InfoTech, Toyota Motor Corporation, 1-6-1 Otemachi, Chiyoda-ku, Tokyo 100-0004, Japan}

\date{August 28, 2026}

\begin{abstract}
Clusters that merge and split have no unique genealogy.  Tracking the host size of a conserved constituent defines $q_*$ by $\langle(S_{t+\tau}/S_t)^{q_*}\rangle=1$, with $q_*\ne0$.  Across coupled map networks, coagulation-fragmentation dynamics, and Vicsek systems, $q_*\simeq\alpha-2$ without fitting, where $c(s)\propto s^{-\alpha}$ on the same size window.  Radius scans locate the known Vicsek interaction radius and recover an independently specified clustering scale in experimental active rods.  Local scale covariance yields the relation, while a tent map counterexample shows when reducing the stationary balance at each parent size to one moment fails.
\end{abstract}

\maketitle

\begin{figure*}[t!]
 \centering
 \includegraphics[width=0.95\textwidth]{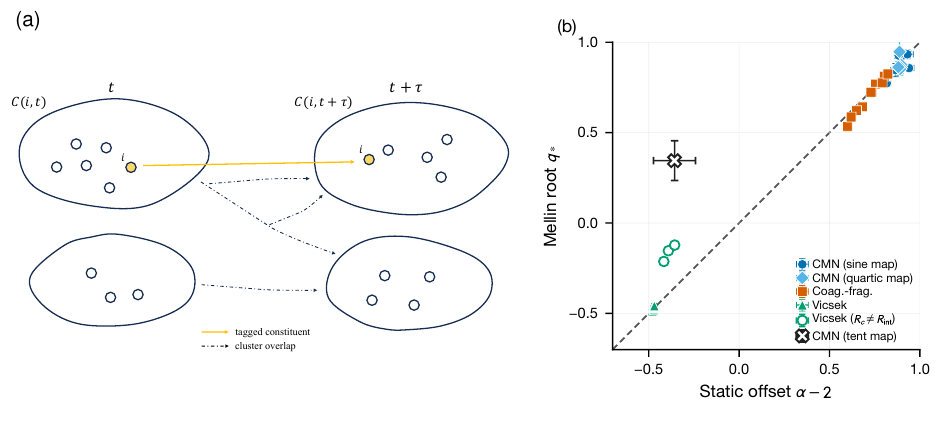}
 \caption{\label{fig:main} (a) Merging and splitting events admit no canonical one-to-one cluster continuation.  The conserved tag $i$ identifies its host clusters $C(i,t)$ and $C(i,t+\tau)$, thereby defining $S_t=|C(i,t)|$ and $S_{t+\tau}=|C(i,t+\tau)|$.  Yellow marks the tag and the solid arrow its continuation; dash-dotted arrows indicate cluster overlap. (b) Mellin root versus static exponent offset for 30 parameter sets (ten per family).  Colors and filled markers distinguish the two CMN maps, coagulation-fragmentation, and Vicsek clusters defined by the physical interaction graph.  Open green circles show reclusterings of the same fixed Vicsek trajectories at the tested values nearest to $R_{\rm int}$ but with $R_c\ne R_{\rm int}$, and the black cross shows the counterexample based on the tent map.  The residual statistics include the 30 filled markers.  The full radius scan is shown in Fig.~\ref{fig:vicsek-graph-scale}.  The dashed line is equality without a fit.  Horizontal and vertical error bars are standard errors of $\alpha-2$ and $q_*$, respectively, across independent realizations; some are smaller than the markers.}
\end{figure*}

Time-resolved tracking now resolves individual trajectories in active colloids, migrating cells, and animal groups~\cite{Ginot2018,Brueckner2021,Katz2011}.  These data raise an inverse problem~\cite{Ballerini2008}: which proximity graph captures the physical collective organization?  A geometrical threshold alone does not test whether the resulting clusters are dynamically consistent.  This ambiguity is acute when clusters merge, split, and exchange constituents, as in coupled chaotic maps, aggregation-fragmentation kinetics, and active matter~\cite{Kaneko1990,ManrubiaMikhailov1999,KrapivskyRednerBenNaim2010,ConnaughtonRajeshZaboronski2004,ConnaughtonRajeshZaboronski2007,HuepeAldana2004,PeruaniDeutschBar2006,PeruaniSchimanskyGeierBar2010,PeruaniBar2013}.  After a merge followed by a split, no canonical one-to-one continuation identifies a descendant cluster.  For observation windows over which constituent identity remains available, we remove this genealogical ambiguity by tagging one constituent and following the size of its host cluster [Fig.~\ref{fig:main}(a)].  This time series of host size supplies a common dynamical observable without reconstructing a genealogy.  In a controlled Vicsek test, its agreement with static scaling selects the known interaction graph among the proximity graphs tested.  In experimental active rods, the same comparison recovers an independently specified clustering scale.  Counting each cluster once supplies the corresponding static descriptor:
\begin{equation}
 c(s)\propto s^{-\alpha}.
 \label{eq:static}
\end{equation}
We compare transfer and static scaling on the same inclusive window of parent sizes, $\Iwin=[s_{\min},s_{\max}]$, where $\alpha$ is the local slope estimated on that interval.

Related studies have characterized the formation and splitting of synchronized clusters in globally coupled maps and coupled map networks~\cite{PopovychMaistrenkoMosekilde2001,JalanAmritkarHu2005}.  Known statistical relations usually exploit the kinetics of a particular system.  Arguments based on constant flux constrain stationary aggregation spectra and correlations that carry the flux~\cite{ConnaughtonRajeshZaboronski2004,ConnaughtonRajeshZaboronski2007}.  Transitions obtained by following particles in a Vicsek-type model can obey detailed balance~\cite{HuepeAldana2004}.  Studies of active clusters connect their size distributions to fusion, fragmentation, particle loss, and geometry~\cite{PeruaniDeutschBar2006,PeruaniSchimanskyGeierBar2010,Theurkauff2012,PeruaniStarruss2012,PeruaniBar2013,Ginot2018}.  Together, these results motivate a common observable: the change in host size seen by a tagged constituent.

An earlier empirical relation for coupled map networks (CMNs) connected cluster statistics to the scaling of the Lyapunov spectrum.  Here we address a different question by measuring the relevant quantities on the same trajectories.  In a chaotic Griffiths phase, the number $N_+(N)$ of positive Lyapunov exponents scales as $N_+(N)\sim N^\beta$.  Shinoda and Kaneko proposed the following relation between $\beta$ and the cluster exponent $\alpha$~\cite{ShinodaKaneko2016}.
\begin{equation}
 \alpha\simeq2(1+\beta).
 \label{eq:sk}
\end{equation}
Equation~\eqref{eq:sk} connects the cluster exponent to the scaling of the Lyapunov spectrum, unlike the kinetic relations above.  Because it was inferred across parameter values, we ask whether both exponents describe the same cluster sizes and time scales on a common trajectory.

The argument proceeds in four steps.  First, sampling a constituent and measuring its host size on a logarithmic scale fix the offset $2$.  Second, we assume local scale covariance: within the chosen window, transition statistics depend primarily on the size ratio rather than on the parent size itself.  At leading order, this reduces the stationary equation with its explicit dependence on parent size to a single moment of the tagged size ratio.  Third, we test the resulting parameter-free relation across three model families.  We then ask whether its residual selects a known interaction graph in controlled active matter and an independently specified clustering scale in experiment.  Finally, a counterexample based on the tent map identifies the failed reduction, and we use the same CMN trajectories to test compatibility with the earlier Lyapunov relation.

Let $\mathcal P_t$ be a partition of $N$ conserved constituents.  Choose a constituent $i$ uniformly and define
\begin{equation}
 S_t=|C(i,t)|,\qquad Y_t=\ln S_t,
 \label{eq:tagged}
\end{equation}
where $C(i,t)$ is the cluster containing $i$.  Such tagged components are familiar in fragmentation-coalescence theory~\cite{Bertoin2006,Berestycki2004,BertoinMartinez2005}.  Here the construction supplies the same observable for deterministic, stochastic, and geometric partitions.

Uniform constituent sampling size-biases the cluster count, $P_{\rm tag}(s)=s c(s)/\sum_u u c(u)$~\cite{Niwa2003,Degond2020}.  Let $r(y)$ denote the corresponding coarse-grained stationary density with respect to $dy$.  Passing from $s$ to $y=\ln s$ contributes the Jacobian $ds/dy=s$, and therefore
\begin{equation}
 r(y)=sP_{\rm tag}(s)\propto s^2c(s)
      \propto e^{-(\alpha-2)y}.
 \label{eq:measure}
\end{equation}
Thus the offset $2$ is fixed: one power comes from constituent sampling and one from the logarithmic size measure.

We call the following leading-order approximation local scale covariance: within the logarithmic window $\ln\Iwin$, the conditional transition statistics depend primarily on the logarithmic size increment $y'-y$, rather than separately on the parent size $y$.  In the interior, the transfer kernel for the tagged size can then be written approximately as $T_\tau(y'|y)=\kappa_\tau(y'-y)$.  Substituting the exponential profile in Eq.~\eqref{eq:measure} into the stationary equation $r(y')=\int dy\,T_\tau(y'|y)r(y)$ and dividing by $r(y')$ gives, with $\xi=y'-y$,
\begin{equation}
 1=\int d\xi\,\kappa_\tau(\xi)e^{(\alpha-2)\xi}.
 \label{eq:balance}
\end{equation}
Since $S_{t+\tau}/S_t=e^\xi$, the corresponding trajectory observable is
\begin{equation}
 \begin{aligned}
 M_\tau(q)&=\left\langle
 \left(\frac{S_{t+\tau}}{S_t}\right)^q
 \middle|S_t\in\Iwin\right\rangle,\\
 M_\tau(\qstar)&=1,\qquad \qstar\ne0.
\end{aligned}
\label{eq:mellin}
\end{equation}
Under the same approximation, $M_\tau(q)\simeq\int d\xi\,\kappa_\tau(\xi)e^{q\xi}$.  Local scale covariance thereby reduces the stationary equation with its explicit dependence on parent size to the scalar condition $M_\tau(q)=1$.  Because $M_\tau(0)=1$ identically, $q=0$ expresses normalization only.  The nonzero root instead reweights growth and shrinkage until their contributions balance; we call it the Mellin root $\qstar$.  Although the average is conditioned on $S_t\in\Iwin$, it includes every child size, even when $S_{t+\tau}\notin\Iwin$.  The window therefore selects only the parent sizes; transitions across its boundaries still contribute to the stationary balance that determines $\qstar$.

Each comparison uses one window of parent sizes and one transfer lag $(\Iwin,\tau)$.  The lag lets cluster membership change while most tagged transitions remain inside the window, and changing it defines a new transfer operator.  Under the same approximation and at fixed $(\Iwin,\tau)$, Eq.~\eqref{eq:balance} implies $M_\tau(\alpha-2)\simeq1$, whereas Eq.~\eqref{eq:mellin} defines $M_\tau(\qstar)=1$.  Therefore,
\begin{equation}
 \qstar\simeq\alpha-2.
 \label{eq:law}
\end{equation}

In stationary dynamics with merging and splitting, Eq.~\eqref{eq:law} has a Mellin form familiar from several settings.  Similar conditions select tails in random affine and multiplicative processes~\cite{Kesten1973,Goldie1991,SornetteCont1997}.  The same Mellin structure also appears in Malthusian identities for tagged fragments and in exponents of count densities in pure fragmentation~\cite{BertoinMartinez2005,KrapivskyBenNaimGrosse2004}.  Here the multiplier is measured directly from the change in host size over a finite lag.  Constituent sampling and the logarithmic size measure fix the offset $2$, so $\qstar$ can be compared with the independently measured local slope $\alpha$ of $c(s)$ without fitting a slope or intercept.

We tested Eq.~\eqref{eq:law} on 30 parameter sets drawn equally from three model families.  In CMNs, proximity in state space defines clusters of maps coupled on a fixed random graph.  The coagulation-fragmentation model partitions a conserved integer mass through stochastic events that merge or split clusters.  Related reversible coagulation-fragmentation chains and uniform split-merge transformations admit explicit invariant measures~\cite{DurrettGranovskyGueron1999,MayerWolfZeitouniZerner2002,DiaconisMayerWolfZeitouniZerner2004}.  In finite, weakly polarized Vicsek states, clusters are connected components of the same moving interaction graph that mediates orientational alignment~\cite{Vicsek1995,Ginelli2016,KyriakopoulosChateGinelli2019,Rahman2026}.  The three families thus reorganize clusters through deterministic chaos, stochastic merging and splitting at fixed total mass, and collective motion, respectively.

For each parameter set, transfer statistics determine the size window and lag in one realization ensemble, while two disjoint ensembles independently estimate $\qstar$ and the slope $\alpha$ of a truncated power-law fit~\cite{Clauset2009,SupplementalMaterial}.  All uncertainties treat the realization as the independent unit.

Figure~\ref{fig:main}(b) shows that the 30 estimates track the parameter-free equality line, with a mean absolute residual of $0.0222$.  The alternatives $\qstar=2(\alpha-2)$ and $\qstar=\alpha-1$ instead give mean absolute residuals of $0.706$ and $1.010$, respectively.  As a separate diagnostic, an orthogonal fit across all three families likewise gives unit slope and zero intercept within uncertainty~\cite{SupplementalMaterial}.

\begin{figure}[t!]
 \centering
 \includegraphics[width=\columnwidth]{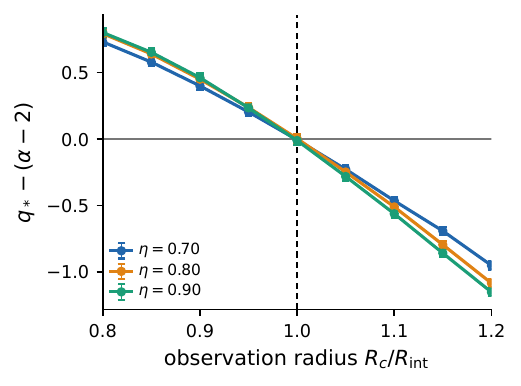}
 \caption{\label{fig:vicsek-graph-scale} Selection of the interaction radius in controlled active matter.  Fixed Vicsek trajectories generated with interaction radius $R_{\rm int}$ are reclustered at $R_c$.  For all three noise values, the residual between the transfer and static exponents is consistent with zero only at the physical graph $R_c/R_{\rm int}=1$ (vertical dashed line); error bars are simultaneous 95\% bootstrap intervals over the radius scan.}
\end{figure}

We next asked whether the relation follows the interaction itself or merely the cluster definition.  Keeping the size window and lag fixed, we reclustered the same Vicsek trajectories at nine observation thresholds, $0.8\le R_c/R_{\rm int}\le1.2$.  The mean absolute residual is $0.0069$ at the physical interaction graph, versus $0.585$ across the 24 points away from that graph.  The simultaneous 95\% bootstrap intervals contain zero only for the three noise values at $R_c=R_{\rm int}$ [Fig.~\ref{fig:vicsek-graph-scale}].  Agreement thus identifies the neighborhood graph that mediates orientational alignment.

\begin{figure*}[t!]
 \centering
 \includegraphics[width=0.90\textwidth]{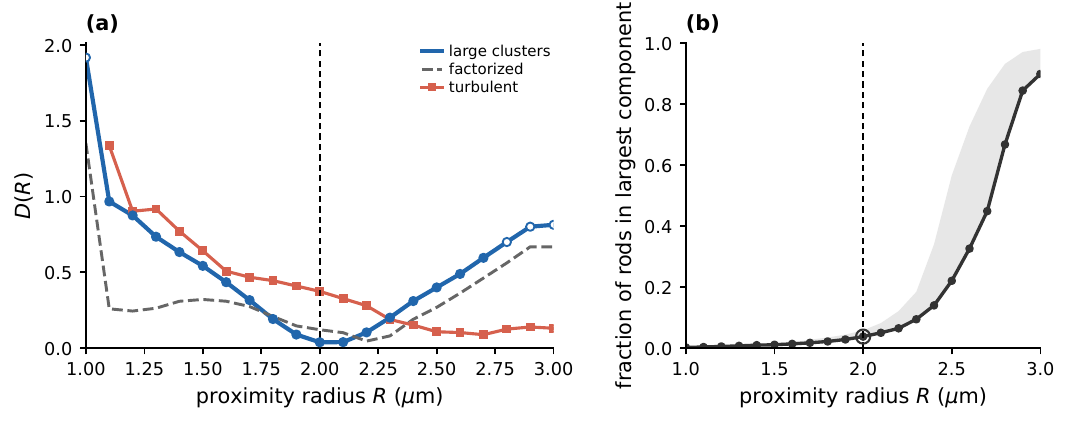}
 \caption{\label{fig:active-rods-scale} Recovery of the clustering radius in experimental active matter.  (a) For the state with large clusters ($\phi=0.725$), the mismatch between temporal blocks $D(R)=\{ |q_{*,A}-(\alpha_B-2)|+|q_{*,B}-(\alpha_A-2)|\}/2$ has a minimum at $2.0$--$2.1\,\mu\mathrm m$, with its grid minimum at the published $2\,\mu\mathrm m$ cutoff.  The minima move to $2.2$ and $2.7\,\mu\mathrm m$ when parent and child sizes are factorized or when the sequence from the turbulent state is used.  Open symbols mark radii outside the fixed count criteria.  (b) The curve is the framewise median fraction of rods in the largest component; the gray band extends from the median to the 95th percentile over retained frames.  The open ring marks the median $0.0376$ at the recovered scale, before the rapid growth at larger $R$.  The vertical dashed lines in both panels mark $2\,\mu\mathrm m$.}
\end{figure*}

Can the same criterion recover a clustering scale specified independently of the relation?  We tested it in publicly deposited trajectories of light-driven active rods~\cite{Shelke2026,ShelkeData2026}.  The source analysis defined clusters with a $2\,\mu\mathrm m$ cutoff.  Holding the size window, lag, and temporal blocks fixed, we varied only the proximity radius.  In the state with large clusters ($\phi=0.725$), $D(R)$ forms a narrow minimum at $2.0$--$2.1\,\mu\mathrm m$, with grid minimum $0.0380$ at the published cutoff [Fig.~\ref{fig:active-rods-scale}(a)].  The minimum shifts when parent and child sizes are factorized and also for the separate sequence in the turbulent state.  At the recovered scale, the framewise median fraction of rods in the largest component is $3.76\%$, placing the residual minimum before the rapid growth of a giant component [Fig.~\ref{fig:active-rods-scale}(b)].  The agreement between the tagged dynamics and static scaling therefore recovers an experimental clustering scale specified independently of the relation.

Where does the reduction to one moment fail?  The CMN with the tent map provides a counterexample: its mean Mellin root and mean $\alpha-2$ have opposite signs [black cross in Fig.~\ref{fig:main}(b)], giving $|\qstar-(\alpha-2)|=0.703$.  To locate the failed step, we retain the dependence on parent size that the single moment averages out.  We apply the measured transition matrix to trial profiles with a power-law form and restore influx from parents outside $\Iwin$.  Minimizing the residual of the stationary equation defines the operator exponent $\Theta_{\rm op}$.  The values of $|\qstar-\Theta_{\rm op}|$ for the 12 diagnostic parameter sets do not overlap those for the tent map.  This locates the failure in reducing the stationary equation with its explicit dependence on parent size to $M_\tau(q)=1$.  Changing the local map continuously shows that this failure can be tuned and is not explained by the dispersion of local slopes alone~\cite{SupplementalMaterial}.

Finally, the earlier Lyapunov relation concerns a trend across parameter values.  We ask a different question: can it account for $\qstar$ on the same CMN trajectories and finite scales?  Combining Eqs.~\eqref{eq:sk} and \eqref{eq:law} predicts $\qstar\simeq2\beta$.  At each $N$, we choose the bin width so that the expected number of other independently sampled nodes in the same bin remains fixed.  We define $\beta_N$ as the slope of $\log N_+$ versus $\log N$ over three system sizes centered on $N$.  At the three intermediate sizes, $\qstar-(\alpha-2)$ ranges from $-0.055$ to $0.027$.  In contrast, all paired 95\% intervals for $(\alpha-2)-2\beta_N$ lie below $-0.50$.  The total number of positive Lyapunov exponents therefore does not account for $\qstar$ at these scales~\cite{SupplementalMaterial}.  The present result therefore shows that Eqs.~\eqref{eq:sk} and \eqref{eq:law} do not provide a common description in this setting.

Tagging a constituent turns an ambiguous cluster genealogy into a well-defined trajectory of its host size.  Across three model families, the Mellin root follows $\qstar\simeq\alpha-2$.  Radius scans localize the known Vicsek interaction graph and, in experimental active rods, recover an independently specified clustering scale.  The counterexample based on the tent map and a diagnostic based on the transition matrix identify the breakdown: one moment no longer captures how the stationary balance depends on parent size.  The comparison on the same CMN trajectories also separates the tagged relation from the total number of positive Lyapunov exponents.  Constituent-tagged transfer thereby connects cluster dynamics to static scaling and provides a criterion for comparing candidate graphs in tracking data.

\begin{acknowledgments}
OpenAI Codex 5.6 Sol assisted with code editing and refactoring, organizing numerical results, discussing their interpretation, and editing the manuscript language.  The author reviewed all AI-assisted code and outputs, independently evaluated all suggestions, verified all analyses, and takes full responsibility for the scientific interpretations, conclusions, and content.
\end{acknowledgments}

\bibliography{references}

\end{document}


\title{Supplemental Material}

\author{Kenji Shinoda}
\email{kenji_shinoda@mail.toyota.co.jp}
\affiliation{InfoTech, Toyota Motor Corporation, 1-6-1 Otemachi, Chiyoda-ku, Tokyo 100-0004, Japan}

\date{}
\maketitle

This Supplemental Material defines the estimators, simulation models, and numerical procedures, and tests sensitivity to the size interval and lag.  It also describes the radius scan for experimental active rods and gives a stationarity calculation that retains the dependence on parent size for the counterexample based on the tent map.

\section{Estimators and numerical models}

Let $n_\tau(s,s')$ count transitions obtained by following a constituent from a cluster of size $s$ to one of size $s'$ after lag $\tau$.  On the inclusive interval of parent sizes $\Iwin=[\ell,h]$, we measure
\begin{align}
 M_\tau(q;\Iwin)
 &=\frac{\displaystyle\sum_{s\in\Iwin}\sum_{s'\geq1}
 n_\tau(s,s')(s'/s)^q}
 {\displaystyle\sum_{s\in\Iwin}\sum_{s'\geq1}n_\tau(s,s')},
 \label{eq:sm-moment}\\
 f_{\rm exit}
 &=\frac{\displaystyle\sum_{s\in\Iwin}\sum_{s'\notin\Iwin}n_\tau(s,s')}
 {\displaystyle\sum_{s\in\Iwin}\sum_{s'\geq1}n_\tau(s,s')}.
 \label{eq:sm-exit}
\end{align}
Only $S_t$ is conditioned to lie in $\Iwin$.  Transitions with $S_{t+\tau}\notin\Iwin$ are still included in the average.  We find the nonzero root of $M_\tau(q;\Iwin)=1$ by a bracketed log-sum-exp search.  In a finite system the moment exists for every finite $q$.  With $\xi=\ln(S_{t+\tau}/S_t)$, strict convexity of $\ln M_\tau(q)$ makes a nonzero root of the ensemble moment unique when the increment distribution is nondegenerate and such a root exists.  Its lag dependence is quantified below.

We fit the independent static ensemble on the same interval with the likelihood for a discrete power law truncated to $\Iwin$:
\begin{equation}
 \mathcal L(\alpha\mid\Iwin)=
 \prod_{s=\ell}^{h}
 \left[\frac{s^{-\alpha}}{\sum_{u=\ell}^{h}u^{-\alpha}}\right]^{c(s)},
 \qquad \Theta\equiv\alpha-2.
 \label{eq:sm-mle}
\end{equation}
We maximize this discrete likelihood directly~\cite{Clauset2009}.  The transfer calculation fixes the interval before the static fit, so $\alpha$ is the local slope on that interval.

\paragraph{Coupled map networks.---}
The dynamics is
\begin{equation}
 x_i(t+1)=(1-\epsilon)f[x_i(t)]+\epsilon\sum_jW_{ij}f[x_j(t)],
 \label{eq:sm-cmn}
\end{equation}
where $W$ is the row-normalized adjacency matrix of an Erd\H{o}s--R\'enyi graph with $N=1024$ and mean degree 20.  The primary grid uses $f(x)=\sin(\pi x)$ over $0.690\le\epsilon\le0.710$ and $f(x)=1-2x^4$ over $0.640\le\epsilon\le0.660$, both in steps of 0.005.  Nodes in the same bin of state space form a cluster.  We calibrate the bin width so that one other independently sampled node is expected in the same bin, $g=1$ [Eq.~\eqref{eq:occupancy}], and use this width for the two measurement blocks.

\paragraph{Coagulation-fragmentation.---}
The process partitions a conserved integer mass $N=2048$.  At each event, two uniformly selected mass units merge their host clusters with probability $p$; otherwise, one selected unit triggers a uniformly cut integer split of its cluster.  The primary grid covers $0.33\le p\le0.42$ in steps of 0.01.  We discard the first $3\times10^5$ events and then record two blocks of $4\times10^5$ events, sampling every 20 events.  We follow 128 passive tags to estimate the transition kernel seen by one constituent.  Because tags within a realization are correlated, the realization, rather than the tag, is the independent unit.  Physical mass remains conserved globally~\cite{KrapivskyRednerBenNaim2010}, while $f_{\rm exit}$ measures tagged probability crossing the chosen size interval.

\paragraph{Vicsek active matter.---}
We simulate $N=512$ particles at density 1 and speed $v_0=0.5$~\cite{Vicsek1995}.  Headings align within $R_{\rm int}=1$ and receive additive angular noise $\eta\pi U[-1,1]$, reduced modulo $2\pi$.  Clusters are connected components of this same instantaneous interaction graph~\cite{Rahman2026}.  The primary grid covers $0.700\le\eta\le0.925$ in steps of 0.025.  We discard the first 1000 steps and then record two 1500-step blocks.  Across this grid, the mean polarization is 0.0396--0.0489, close to the finite-size value $\sqrt{\pi}/(2\sqrt N)=0.03917$ expected for $N$ independent headings; these are weakly polarized finite systems.

Each realization contains two nonoverlapping blocks, whose estimates are averaged within the realization.  Realizations shared across parameter points within a model family are resampled jointly, whereas different families are treated independently.  The realization is the resampling unit; its particle times, passive tags, and two blocks remain grouped.

\paragraph{Ensembles and uncertainty.---}
The primary comparison comprises 30 parameter sets, ten from each model family.  Eight realizations determine each $(\Iwin,\tau)$ pair from transfer statistics; separate ensembles of eight and five realizations determine $\qstar$ and $\alpha-2$, respectively.

The detailed diagnostics use 12 parameter sets: two CMN parameter sets using the sine map, two using the quartic map, five coagulation-fragmentation probabilities, and three Vicsek noise values.  This set supplies the window and lag selection, their sensitivity checks, and the stationarity diagnostics below that retain the dependence on parent size.

An admissible pair $(\Iwin,\tau)$ has a nonzero root, at least $10^4$ retained transitions per block, a common root sign, and stable roots within and between realizations and between the two time blocks.  For the stability checks, let $m=\max(|\operatorname{median}(\qstar)|,0.25)$, with the median taken over all roots from all realizations and blocks.  Writing MAD for the median absolute deviation, we require $1.4826\operatorname{MAD}(\qstar)/m\le0.25$, $\operatorname{SD}(\overline q_{*,r})/m\le0.20$ over realization means, and $\langle|q_{*,r,A}-q_{*,r,B}|\rangle_r/m\le0.25$.

An admissible static block has at least 500 observations, six occupied sizes, and discrete Kolmogorov--Smirnov distance at most 0.12.  The largest observed static distances are 0.02401 in the 30-point comparison and 0.0231 for the 12 diagnostic parameter sets.

Bootstrap resampling uses the realization as the independent unit.  Within each family, common realization indices are resampled jointly across its ten parameter points, whereas the disjoint transfer and static ensembles are resampled independently.  A familywise simultaneous half-width is the 95th percentile of the maximum centered residual change across the ten points.  Bootstrap intervals are evaluated with the selected pair $(\Iwin,\tau)$ held fixed.

Across the 30 parameter points, the signed mean residual is $-0.0098$, with a 95\% bootstrap interval $[-0.0175,-0.0009]$.

We summarize the 30 estimates by unweighted total least squares using orthogonal distances.  The resulting relation $\qstar=a+b(\alpha-2)$ has $b=0.9873$ and $a=-0.0050$.  A $20\,000$-replicate bootstrap preserving the familywise realization dependence gives 95\% intervals $[0.9768,1.0004]$ for $b$ and $[-0.0098,0.0001]$ for $a$, including $b=1$ and $a=0$.  The bootstrap intervals for the increases in mean absolute residual are $[0.6669,0.6895]$ for $\qstar=2(\alpha-2)$ and $[0.9723,0.9922]$ for $\qstar=\alpha-1$, relative to the parameter-free relation.

\subsection{Controls and CMN resolution}

\paragraph{Vicsek radius control.---}
We hold the Vicsek trajectories at $\eta=0.70,0.80,0.90$ fixed and reconstruct the observed graph at each radius in $0.80\le R_c/R_{\rm int}\le1.20$, in steps of 0.05.  Clusters are the connected components of each graph.  We retain separate ensembles of eight realizations for the transfer statistics and five for the static statistics, together with $\Iwin=[5,20]$ and $\tau=4$.  Realizations are resampled jointly across the radius grid.  Varying the observation radius tests whether the agreement localizes at the physical radius $R_c=R_{\rm int}$.  Lines connect adjacent radii.

\paragraph{Radius scan for experimental active rods.---}

We reanalyzed published trajectories of light-driven TiO$_2$--SiO$_2$ rods with aspect ratio 7.5~\cite{Shelke2026,ShelkeData2026}.  The two deposited conditions have nominal particle numbers and area fractions $(N,\phi)=(2832,0.725)$ and $(1532,0.3928)$; the source study describes the former as a state with large clusters and the latter as a turbulent state.  The images were acquired at 20 frames per second, and the deposited coordinates use $0.0696\,\mu\mathrm m$ per pixel.  Each retained transition follows the same identified rod between its two observed endpoints.

The source analysis used DBSCAN~\cite{EsterKriegelSanderXu1996} with radius $2\,\mu\mathrm m$ and a core threshold of two points.  For clusters of size at least two, DBSCAN therefore returns the connected components of a Euclidean radius graph constructed from the deposited positions of the rod heads.  We treat DBSCAN noise points as singleton components, completing the partition of observed particles; the analyzed interval of parent sizes begins at $s=4$.

We held $\Iwin=[4,16]$ and $\tau=10$ frames $=0.5\,\mathrm s$ fixed while varying only the graph radius over $1.0\le R\le3.0\,\mu\mathrm m$ in steps of $0.1\,\mu\mathrm m$.  For the state with large clusters, blocks $A$ and $B$ comprise frames 70--139 and 140--209.  For the turbulent state, they comprise frames 116--182 and 183--249.  We count each static cluster once per frame.  At each source frame, a rod observed $\tau$ frames later contributes a transition if its parent cluster has $s\in\Iwin$; all child sizes are included.

To compare the blocks in both directions, we use
\begin{equation}
 D(R)=\frac{1}{2}\left[
 \left|q_{*,A}(R)-\bigl(\alpha_B(R)-2\bigr)\right|
 +\left|q_{*,B}(R)-\bigl(\alpha_A(R)-2\bigr)\right|
 \right].
 \label{eq:sm-rod-residual}
\end{equation}
In Fig.~3(a) of the Letter, filled points denote radii for which both nonzero roots and both static fits are identifiable.  Each block also supplies at least $10^4$ such transitions, and each static fit contains at least 500 cluster observations in $\Iwin$.

At $R=2\,\mu\mathrm m$, blocks $A$ and $B$ give $q_*=-0.4169$ and $-0.4075$, while their static estimates give $\alpha-2=-0.4485$ and $-0.4519$.  The corresponding mismatches between blocks are 0.0350 and 0.0411, whose mean gives $D(2\,\mu\mathrm m)=0.0380$.  The blocks contain 49,275 and 47,504 retained transitions.  Among rods whose parent cluster lies in $\Iwin$, the fractions observed again after $\tau$ are 0.923 and 0.919.  Their static fits contain 8,142 and 7,919 observations, with discrete Kolmogorov--Smirnov distances 0.0150 and 0.0142.  Across the same scan, these distances are minimized at $R=1.6\,\mu\mathrm m$ in block $A$ and $R=1.9\,\mu\mathrm m$ in block $B$, rather than at the residual minimum at $2.0$--$2.1\,\mu\mathrm m$.  For the structural diagnostic in the Letter, we compute the fraction of rods in the largest component in each frame.  We plot its median and 95th percentile over frames.

Two comparisons test the specificity of the recovered scale.  First, we factorized the selected transition counts into the product of their parent and child marginals, preserving both marginals while removing their dependence.  Its mismatch is 0.1209 at $R=2\,\mu\mathrm m$, and its minimum shifts to $2.2\,\mu\mathrm m$, where $D(2.2\,\mu\mathrm m)=0.0458$.  Second, the sequence in the turbulent state gives $D(2\,\mu\mathrm m)=0.3742$ and reaches its minimum at $2.7\,\mu\mathrm m$.

At $R=2\,\mu\mathrm m$, the roots computed separately for each parent size range from $-0.891$ to $-0.049$ in block $A$ and from $-0.963$ to $-0.136$ in block $B$.  The reported agreement is the aggregate moment over $\Iwin=[4,16]$, where the fitted $\alpha<2$ is interpreted as a local slope.  The recovered value is the clustering scale specified independently in the source analysis.

\paragraph{CMN resolution and Lyapunov compatibility.---}
In a CMN, nodes occupying the same bin of width $\delta$ in state space are assigned to the same cluster.  The observation therefore has a resolution scale, and a fixed threshold can create a preasymptotic scaling regime distinct from the underlying asymptote~\cite{ShinodaKaneko2016,FontClos2015}.  To quantify the contribution expected even for independent nodes, we define
\begin{equation}
 g(N,\delta)=(N-1)\sum_b p_b(\delta)^2,
 \label{eq:occupancy}
\end{equation}
where $p_b(\delta)$ is the probability that one node occupies bin $b$.  Thus $g(N,\delta)$ is the expected number of other independently sampled nodes in the focal node's bin.  Holding $g$ fixed keeps this baseline constant as $N$ changes.

For this test, the width $\delta(N)$ is calibrated to keep $g$ fixed.  We use $N=512$, 1024, 2048, 4096, and 8192.  Exact cluster counts, tagged transitions, and the Lyapunov spectrum are then measured over the same 32000-step interval of each trajectory, using four graph realizations paired across sizes.  The spectrum is computed with the Benettin method using reduced QR factorizations~\cite{BenettinGalganiGiorgilliStrelcyn1980a,BenettinGalganiGiorgilliStrelcyn1980b}.  We jointly resample entire graph realizations and pool the reported cluster estimates over four offsets of the bin grid in state space.

At $\epsilon=0.54$, we use $(\Iwin,\tau)=([12,48],1)$ for $N=512$--8192.  Writing $\beta_N$ for the slope of $\log N_+$ versus $\log N$ over the three sizes $N/2$, $N$, and $2N$, the estimates obtained at fixed $g$ for $N=1024$, 2048, and 4096 give $\alpha-2=0.888$--$0.902$ and $2\beta_N=1.518$--$1.810$.  The separation remains under fixed resolution, alternative sign counts, size windows and estimators, and in a second ensemble.

\section{Sensitivity to size window and lag}

\subsection{Choice of size window and lag}

The CMN map families and Vicsek use one common choice of $(\Iwin,\tau)$ per family.  Because coagulation-fragmentation relaxes strongly with $p$, its candidates span $\tau=20$--160 events in factors of 2 and $\Iwin=[\ell,4\ell]$.  The ranking score favors a nonzero root well separated from zero, together with exchange across the interval boundary that is measurable but not dominant:
\begin{equation}
 \mathcal S=\left(\frac{\qstar-0.70}{0.30}\right)^2+
   \left(\frac{f_{\rm exit}-0.14}{0.03}\right)^2
 \label{eq:sm-selector}
\end{equation}
after imposing the common admissibility criteria.

Table~\ref{tab:selection-sensitivity} shows the variation in residuals across 135 choices of the centers and relative weights in Eq.~\eqref{eq:sm-selector}.

\begin{table}[t!]
\caption{Sensitivity to the parameters of the selection rule in Eq.~\eqref{eq:sm-selector}.  Five choices for the root center, three for the center of $f_{\rm exit}$, and nine for their relative weights define 135 parameter choices and yield 65 distinct selections of five $(\Iwin,\tau)$ pairs.  The columns summarize the distribution over these 135 choices after averaging over either all 12 diagnostic parameter sets or the five coagulation-fragmentation sets whose pairs vary.}
\label{tab:selection-sensitivity}
\centering
\begin{ruledtabular}
\begin{tabular}{lccc}
Scope & Median & 5th--95th percentile & Maximum\\
\hline
12 diagnostic parameter sets & 0.0402 & 0.0168--0.0654 & 0.0756\\
Five coagulation-fragmentation sets & 0.0736 & 0.0176--0.1340 & 0.1586\\
\end{tabular}
\end{ruledtabular}
\end{table}

The static slope is most sensitive near the lower boundary of the size interval selected from transfer data.  Table~\ref{tab:window} therefore recomputes both sides after common boundary perturbations.

\begin{table}[t!]
\caption{Sensitivity to interval boundaries for the 12 diagnostic parameter sets.  Each paired entry gives the $-1$ and $+1$ boundary perturbations, in that order.  $\Delta\Theta$ is measured from the reference fit.  The last two columns recompute both sides on the perturbed common interval.}
\label{tab:window}
\centering
\begin{ruledtabular}
\begin{tabular}{lcccc}
Change & median $|\Delta\Theta|$ & max. $|\Delta\Theta|$ & mean abs. residual & max. residual\\
\hline
none & 0 & 0 & 0.0214 & 0.0563\\
lower boundary $-1/+1$ & 0.0526/0.0603 & 0.1732/0.1388 & 0.0739/0.0493 & 0.1990/0.1773\\
upper boundary $-1/+1$ & 0.0073/0.0065 & 0.0164/0.0157 & 0.0255/0.0232 & 0.0627/0.0515\\
both boundaries $-1/+1$ & 0.0579/0.0649 & 0.1855/0.1497 & 0.0845/0.0570 & 0.2109/0.1887\\
\end{tabular}
\end{ruledtabular}
\end{table}

The lower boundary marks entry into the local scaling regime, whereas the upper boundary lies within it.  The lower boundary selected from transfer data therefore also fixes the static fitting interval.  Static time stationarity is less sensitive: averaged over parameter sets, the signed A--B drift in $\alpha-2$ is 0.0042, the mean absolute drift is 0.0233, and the largest absolute drift is 0.0544.

Table~\ref{tab:lag-sensitivity} reports the mean absolute residual, rather than $\qstar$ itself, at each lag.  Because local scale covariance is a finite-window approximation here, the variation with lag provides a sensitivity check on the range over which the single-moment reduction remains adequate.  At the lag fixed for each family or selected from transfer statistics, the mean absolute residual over the two ordered block comparisons is 0.044 overall.

\begin{table}[t!]
\caption{Mean absolute residual by family for the 12 diagnostic parameter sets at each lag measured in recorded samples.  For coagulation-fragmentation, lags of 1, 2, 4, and 8 recorded samples correspond to 20, 40, 80, and 160 events, respectively.}
\label{tab:lag-sensitivity}
\centering
\begin{ruledtabular}
\begin{tabular}{lcccc}
 & \multicolumn{4}{c}{Lag (recorded samples)}\\
Family & 1 & 2 & 4 & 8\\
\hline
CMN & 0.136 & 0.329 & 0.056 & 0.062\\
Coagulation-fragmentation & 0.104 & 0.079 & 0.060 & 0.089\\
Vicsek & 0.059 & 0.027 & 0.007 & 0.028\\
\end{tabular}
\end{ruledtabular}
\end{table}

For the five selected coagulation-fragmentation parameter sets, the raw exit fraction ranges from 0.125 to 0.163.  Outgoing transitions are dynamically important: omitting child sizes outside $\Iwin$ shifts $\qstar$ by 0.480 on average across 80 estimates, one for each block in each realization.

For the Vicsek parameter sets, $\alpha<2$, so the size-biased power law is controlled by the finite-$N$ cutoff.  We therefore compare the normalized distribution of sizes seen by a tagged particle, conditioned on $s\in[5,20]$.

\section{Stationarity at each parent size: the counterexample based on the tent map}

Let $P_{II}$ be the measured matrix for transitions between sizes inside $\Iwin$, and let $b_I$ be the tagged probability weight entering $\Iwin$ from parents outside it.  For
\begin{equation}
 v_\Theta(s)=A s^{-(\Theta+1)},\qquad s\in\Iwin,
 \label{eq:sm-trial}
\end{equation}
$A$ is chosen so that $V=\sum_{s\in\Iwin}v_\Theta(s)$ equals the empirical tagged probability weight in the interval.  The exponent $\Thetaop$ minimizes
\begin{equation}
 \chi^2(\Theta)=\frac{1}{V}\sum_{s\in\Iwin}
 \frac{\bigl([v_\Theta P_{II}+b_I]_s-v_\Theta(s)\bigr)^2}
 {\max[v_\Theta(s),10^{-12}V]},
 \qquad
 \Thetaop=\underset{-1.25\leq\Theta\leq2.25}{\arg\min}\,\chi^2(\Theta).
 \label{eq:sm-operator}
\end{equation}
The source block supplies $P_{II}$ and $b_I$.  The other block supplies the static comparison.  We also repeat the calculation with $b_I$ omitted.

\begin{figure}[b!]
\centering
\includegraphics[width=0.92\linewidth]{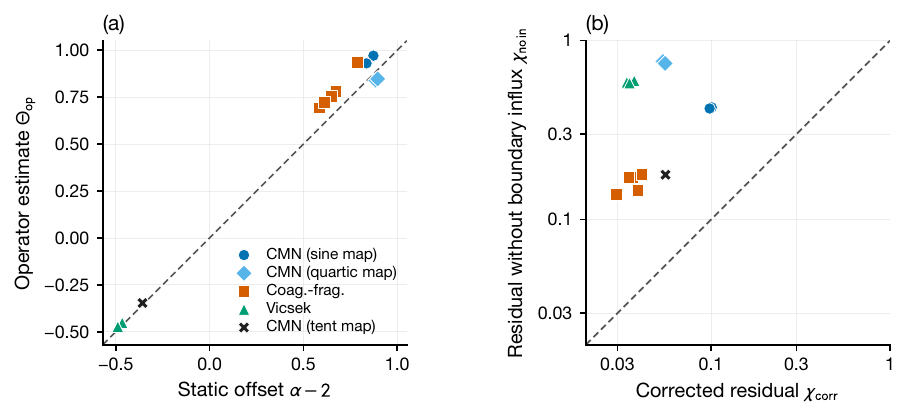}
 \caption{\label{fig:sm-operator} (a) Exponent obtained with incoming tagged probability weight restored, plotted against the static exponent from the other block for the 12 diagnostic parameter sets and the counterexample based on the tent map.  (b) Residual of the stationary equation at the static exponent from the other block when the incoming tagged probability weight is omitted, plotted against the residual with that weight restored.  Each point is the mean for one parameter set (13 in total); colors and markers are common to both panels and to Fig.~1(b) of the Letter.  Dashed lines are equality in data coordinates.  Both axes in panel (b) are logarithmic.  The black cross denotes the CMN with the tent map, the counterexample to the relation $q_*=\alpha-2$.}
\end{figure}

In Fig.~\ref{fig:sm-operator}(b), $\chi_{\rm corr}$ is the square root of Eq.~\eqref{eq:sm-operator} evaluated at the static exponent from the other block.  The quantity $\chi_{\rm no\,in}$ is the corresponding residual with $b_I$ omitted.

The scalar condition $M_\tau(q)=1$ averages over parent size, whereas the matrix calculation retains this dependence.  The 12 diagnostic parameter sets yield 120 comparisons, with each block used in turn as the source.  For the Mellin root, the mean absolute residual against the static exponent from the other block is 0.0442.  The corresponding values are 0.0790 for the matrix calculation with incoming boundary weight restored and 0.3437 with that weight omitted.  The matrix calculation therefore tests the single-moment reduction without averaging over parent size.

For five realizations of the CMN with the tent map, the mean Mellin root is 0.345 and the mean $\alpha-2$ is $-0.358$, giving $|\qstar-(\alpha-2)|=0.703$.  The matrix calculation has mean absolute residual 0.162 without restored influx and 0.120 with it.  The incoming and outgoing tagged probability fractions are both 0.0459, within the 0.0252--0.1982 range of the 12 diagnostic parameter sets.  Thus the magnitude of the flux across the interval boundary does not distinguish the tent map.

\FloatBarrier
\subsection{Controlled deformation}

To probe the failure for the tent map, we consider the following family of full-branch maps:
\begin{equation}
 f_a(x)=1-2|x|+a|x|(1-|x|),\qquad
 0\le a\le2.0\quad\text{in steps of }0.4.
 \label{eq:sm-slope-family}
\end{equation}
Here $a=0$ is the exact tent map, every $a<2$ retains its sharp kink, and $a=2$ is the smooth map $1-2x^2$.  In an independent ensemble, the standard deviation of $\ln|f'_a|$ increases monotonically from 0 at $a=0$ to 0.910 at $a=2$.
For each graph realization, $\epsilon_a$ is chosen so that the largest mean Lyapunov exponent transverse to synchronization~\cite{PecoraCarroll1998} equals its value for the reference tent map:
\begin{equation}
 \lambda_{\rm loc}(a)+\ln r_W(\epsilon_a)
 =\ln2+\ln r_W(0.85),\qquad
 r_W(\epsilon)=\max_{k>1}|1-\epsilon+\epsilon\lambda_k(W)|.
 \label{eq:sm-transverse-match}
\end{equation}
Here $\lambda_{\rm loc}(a)$ is the mean of $\ln|f'_a|$ for the isolated map, and $\lambda_k(W)$ are the eigenvalues of $W$, with $k>1$ indexing modes transverse to synchronization.  Equation~\eqref{eq:sm-transverse-match} holds the largest mean transverse Lyapunov exponent at its reference value while allowing the remaining transverse spectrum and asynchronous invariant dynamics to vary.  All other parameters equal those of the reference calculation for the tent map: $N=1024$, mean degree 20, $g=1$, and $\Iwin=[12,48]$ at lag one.  We discard the first 20000 steps, perform a separate 1000-step resolution calibration, and then record two 8000-step blocks.  For each choice of source block, that block supplies $\qstar$ and $\Thetaop$, while the other block supplies the estimate $\Theta_{\rm stat}$ from the truncated power-law fit.

At all six values of $a$, we measure $\Delta_a=|\qstar-\Thetaop|$, averaging over the two choices of source block within each realization.  The mean paired reduction $\Delta_0-\Delta_{1.6}$ is 0.528, with 95\% bootstrap interval $[0.501,0.556]$, and $\Delta_a$ decreases in all ten matched realizations.  At $a=2$, however, it rises again while the dispersion of local log slopes continues to increase.

\bibliography{references}